\documentclass[10pt, twoside]{amsart}

\newcommand{\Papersize}{a4paper} 

\usepackage{amssymb}
\usepackage{amsmath}
\usepackage{amsthm}
\usepackage{amscd}
\usepackage{graphicx}
\usepackage{floatflt}
\usepackage{longtable}
\usepackage[ansinew]{inputenc}
\usepackage{ifpdf}
\usepackage{ifthen}
\usepackage{dsfont}
\usepackage{cite}
\usepackage{bbm}

\ifthenelse{\equal{\Papersize}{a4paper}}{\usepackage[dvips,includeall,papersize={210mm,
297mm},marginparsep=0cm,marginparwidth=0cm,left=2.5cm,textwidth=16cm,top=2cm,bottom=2cm]{geometry}}{}

\ifthenelse{\equal{\Papersize}{letterpaper}}{\usepackage[dvips,includeall,paper=letterpaper,marginparsep=0cm,marginparwidth=0cm,left=2.795cm,textwidth=16cm,top=2cm,bottom=2cm]{geometry}}{}

\newcommand{\ro}{\mathrm}

\newcommand{\eps}{\varepsilon}
\renewcommand{\phi}{\varphi}
\renewcommand{\hat}{\widehat}
\renewcommand{\tilde}{\widetilde}

\newcommand{\nhalf}[1][N]{\frac{#1}{2}}

\renewcommand{\Im}{\ro{Im}}

\newcommand{\Htild}{\tilde{H}}

\newcommand{\nablal}{\nabla_l}

\newcommand{\Deltal}{\Delta_l}

\newcommand{\mathbbppsi}[1][\psi]{\mathbb{P}^{#1}}
\newcommand{\PP}[1][\psi]{\mathbb{P}^{#1}}

\newcommand{\tex}[1][B_{l,R}]{t^{#1}_{\text{ex}}}
\newcommand{\texone}[1][(\bx_0)]{\tex[B_{1,R}]#1}
\newcommand{\textwo}[1][(\bx_0)]{\tex[B_{2,R}]#1}

\newcommand{\massl}{m_l}

\newcommand{\bx}{\boldsymbol x}
\newcommand{\bxl}{\boldsymbol x_l}
\newcommand{\bgx}{\boldsymbol X}
\newcommand{\bgxone}{{\boldsymbol X}_1}
\newcommand{\bgxN}{{\boldsymbol X}_N}
\newcommand{\bgxpsi}{{\boldsymbol X}^{\psi}}

\newcommand{\bbqqt}[1][t]{{\boldsymbol X}^{\psi}(\bx_0,#1)}
\newcommand{\bbqqtl}[1][t]{{\boldsymbol X}_l^{\psi}(\bx_0,#1)}
\newcommand{\bbqqtone}[1][t]{{\boldsymbol X}_1^{\psi}(\bx_0,#1)}
\newcommand{\bbqqttwo}[1][t]{{\boldsymbol X}_2^{\psi}(\bx_0,#1)}
\newcommand{\bbqqtN}[1][t]{{\boldsymbol X}_N^{\psi}(\bx_0,#1)}

\newcommand{\bgy}{\boldsymbol Y}
\newcommand{\bz}{\boldsymbol z}
\newcommand{\bgz}{\boldsymbol Z}

\newcommand{\TR}[1][(\bx_0)]{T^R#1}

\newcommand{\flow}[1][T,0]{\Phi^\psi_{#1}}

\newcommand{\psixt}[1][\bgx_1,T]{\psi_{#1}}
\newcommand{\psix}[1][\bgx_1]{\psi_{#1}}

\newcommand{\bvinfty}{\bv^{\psi}_{\infty}}
\newcommand{\vinfty}{v^{\psi}_{\infty}}

\newcommand{\bvlinfty}{\bv^{\psi}_{\infty,l}}

\newcommand{\vlinfty}[1][l]{v^{\psi}_{\infty,#1}}

\newcommand{\pouth}{\hat{\psi}_{\text{out}}}

\newcommand{\R}{\mathbb{R}}

\newcommand{\N}{\mathbb{N}}
\newcommand{\Rn}[1][3N]{\mathbb{R}^{#1}}

\newcommand{\ltrN}[1][3N]{L^2(\Rn[#1])}

\newcommand{\g}{\mathcal{G}}

\newcommand{\G}{\mathcal{G}}

\newcommand{\V}[1][RTC]{\mathcal{V}_{#1}}

\newcommand{\bk}{\boldsymbol k}
\newcommand{\bj}{\boldsymbol j}
\newcommand{\bsig}{\boldsymbol \sigma}
\newcommand{\bv}{\boldsymbol v}

\newcommand{\bp}{\boldsymbol p}

\theoremstyle{plain}
\newtheorem{thm}{Theorem}
\newcommand{\bthe}{\begin{thm}}
\newcommand{\ethe}{\end{thm}}

\newtheorem{lem}{Lemma}
\newcommand{\blem}{\begin{lem}}
\newcommand{\elem}{\end{lem}}

\newtheorem{cor}{Corollary}
\newcommand{\bcor}{\begin{cor}}
\newcommand{\ecor}{\end{cor}}

\newtheorem{defi}{Definition}
\newcommand{\bde}{\begin{defi}}
\newcommand{\ede}{\end{defi}}

\newtheorem{prop}{Proposition}
\newcommand{\bprop}{\begin{prop}}
\newcommand{\eprop}{\end{prop}}

\newtheorem{conj}{Conjecture}
\newcommand{\bconj}{\begin{conj}}
\newcommand{\econj}{\end{conj}}

\theoremstyle{definition}
\newtheorem{rem}{Remark}
\newcommand{\brem}{\begin{rem}}
\newcommand{\erem}{\end{rem}}

\newenvironment{pro}[1][\negthickspace]{\noindent{\bf Proof #1.\:}}{\hfill $\Box$}
\newcommand{\bpro}{\begin{pro}}
\newcommand{\epro}{\end{pro}}

\newcounter{cond}
\newenvironment{cond}[0]{\refstepcounter{cond}
\begin{itemize}\item[\bfseries{A\arabic{cond}.}]}{\end{itemize}}
\newcommand{\bcond}{\begin{cond}}
\newcommand{\econd}{\end{cond}}

\begin{document}
\title[On the Quantum Mechanical Scattering Statistics of Many Particles]{\bf On the Quantum Mechanical Scattering Statistics\\ of Many Particles }

\author[D.\ D\"urr]{Detlef D\"urr }
\author[M.\ Kolb]{Martin Kolb}
\author[T.\ Moser]{Tilo Moser}
\author[S.\ R\"omer]{Sarah R\"omer}

\address{Mathematisches Institut der LMU,
Theresienstr. 39, 80333 M\"unchen, Germany}
\email{duerr@math.lmu.de, kolb@math.lmu.de, moser@math.lmu.de, roemer@math.lmu.de}
\parindent=0pt

\subjclass[2010]{Primary  81U10; Secondary 81U20}
\keywords{$N$-body potential scattering, detection probabilities, Bohmian mechanics}

\thispagestyle{empty}

\begin{abstract}
The probability of a quantum particle being detected in a given solid angle is determined by the $S$-matrix. The explanation of this fact in time dependent scattering theory is often linked to the quantum flux, since the quantum flux integrated against a (detector-) surface and over a time interval can be viewed as the probability that the particle crosses this surface within the given time interval. Regarding many particle scattering, however, this argument is no longer valid, as each particle arrives at the detector at its own random time. While various treatments of this problem can be envisaged, here we present a straightforward Bohmian analysis of many particle potential scattering from which the $S$-matrix probability emerges in the limit of large distances.
\end{abstract}
\maketitle

\section{Introduction}\label{chintro}
In a scattering experiment the central quantity is the cross section, whose derivation is based on the probability that particles are detected in certain solid angles. This probability is computed from the $S$-matrix and is given by
\begin{equation}
\label{1}
\int\limits_{C_{\Sigma_1}}d^3k_1\ldots\int\limits_{C_{\Sigma_N}}d^3k_N\left|\pouth(\bk)\right|^2
\end{equation}
where $\pouth(\bk)$ is the Fourier transform of the $N$ particle scattering state's outgoing asymptote and the $C_{\Sigma_k}$s are the cones (in momentum space) subtending the solid angles in question.

It is a common argument (see e.g.\ \cite{cohen2:97}) that the $S$-matrix formalism can be based on the quantum flux $\bj^{\psi}$ given by
\begin{equation*}
\bj^\psi(\bx,t)=\Im\left(\psi^*(\bx,t)\nabla\psi(\bx,t)\right)\,.
\end{equation*}
According to ordinary statistical mechanics $|\bj^{\psi}(\bx,t)\cdot d\bsig| dt$ gives the probability that a particle crosses the surface element $d\bsig$ during time $dt$. In \cite{combes:75} the connection between quantum flux and \eqref{1} was posed as a theorem, the so called flux-across-surfaces theorem. For one particle potential scattering it asserts that (see \cite{amrein1:97,amrein2:97,duerr1:04,teufel2:99} for various versions)
\begin{equation*}
\lim\limits_{R\to\infty}\int\limits_{0}^\infty\int\limits_{R\Sigma}\bj^{\psi}(\bx,t)\cdot d\bsig dt
	=\lim\limits_{R\to\infty}\int\limits_{0}^\infty\int\limits_{R\Sigma}|\bj^{\psi}(\bx,t)\cdot d\bsig|dt
	=\int\limits_{C_{\Sigma}}\left|\pouth(\bk)\right|^2d^3k\,.
\end{equation*}
Here $R$ is the ``large'' radius of a ball centered around the scattering center and $\Sigma$ is a solid angle. $R\Sigma$ stands for the piece of the ball's spherical surface that subtends $\Sigma$ and $ d\bsig $ for the positively oriented infinitesimal surface element. Asymptotically, the flux $\bj^{\psi}$ points outwards and in that case the flux integrated against the (detector) surface $R\Sigma$ and over all times  is the probability that the particle crosses that (detector) surface at some time (see e.g.\ \cite{daumer:94,daumer:96,daumerflux2:97,duerrteufel:09}). When more than one particle is scattered, the role of the quantum flux is less clear \cite{duerrteufel:02}. Indeed, the crossing times are random, i.e.\ if more than one particle is scattered, one has a different crossing time for \emph{each} particle and in such a setting the single-time many particle flux is of little relevance. Put in simple terms, what one wishes to compute is the joint probability

\begin{equation}
\label{eq.Pcross}
\mathbb{P}\left(\text{particle one crosses }R\Sigma_1, \text{particle two crosses }R\Sigma_2,..\right),
\end{equation}
which in the limit $R \to \infty$ should become \eqref{1}.

\begin{figure}[h]
\begin{center}
\includegraphics[width=8.5cm]{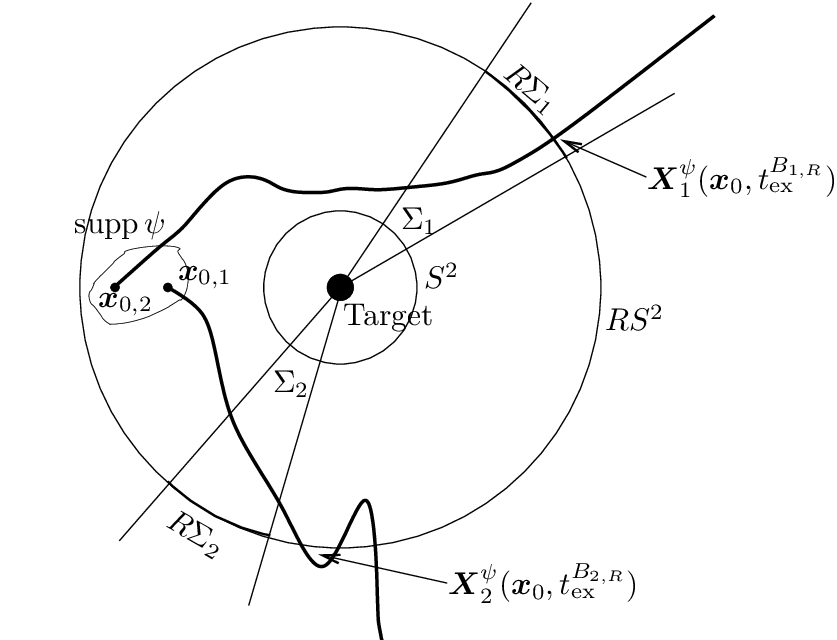}
\end{center}
\caption{Sketch of the scattering situation for $N=2$. }\label{fig.scatt-sit}
\end{figure}

In Bohmian mechanics the joint probability \eqref{eq.Pcross} is readily defined, where  $\mathbb{P}$ is the measure $\mathbbppsi$ on the particles' initial positions with density  $|\psi|^2$. By virtue of the continuity of Bohmian trajectories, first crossing times through boundaries of regions in space are well defined. Let $\tex$  be the first exit time of the $l$th particle from  the ball $x_l<R$  then $\bbqqtl[\tex]\in\Rn[3]$ is the position of the $l$th particle at this time (see Figure \ref{fig.scatt-sit}). We shall show that
\begin{equation}
\label{eq.crossing-prop}
\lim\limits_{R\to\infty}\mathbbppsi\left(\bbqqtl[\tex]\in R\Sigma_l\quad\forall\, l\in\{1,\ldots,N\}\right)
        =\int\limits_{C_{\Sigma_1}}\hspace{-0.2cm}\;
        \ldots\negthickspace\int\limits_{C_{\Sigma_N}}\hspace{-0.2cm}\;\left|\pouth(\bk)\right|^2d^3k_1\cdots d^3k_N\,.
\end{equation}

The idea for proving \eqref{eq.crossing-prop} is rather simple: Far away from the scattering center, the particles' trajectories become more or less straight lines directed along the asymptotic velocity
$$
\bvinfty(\bx_0):=\lim\limits_{t\to\infty}\frac{{\bgx}^\psi(\bx_0,t)}{t}\,,
$$
which is $|\pouth|^2$-distributed\footnote{We use natural units where $\bp=\bk=\bv.$}.

The asymptotic straight line motion of the Bohmian trajectories arises because for large times the wave function is close to a local plane wave
\begin{equation}
\label{eq.loc-pl-w}
e^{-iHt}\psi \approx (it)^{-\frac{3N}{2}}e^{i\frac{x^2}{2t}}\widehat{\psi}_{\text{out}}\left(\frac{\bx}{t}\right)\,,
\end{equation}
where this approximation needs, however, to hold in a sense stronger than $L^2$. Establishing this is part of the technical work that forces us to consider only noninteracting -- but nevertheless entangled -- particles that are scattered off a fixed potential, like e.g.\ EPR pairs. Moreover, since the Bohmian velocity field, just as the flux,  involves derivatives of the wave function, we shall need also stronger requirements on the potential than what one is accustomed to in $S$-Matrix theory.

A realistic description of the scattering process, where detection takes place at finite but random times has to take into account that the detection of one particle leads to a ``collapse of the entangled wave function'' which may
influence the detection statistics of the remaining particles.
Hence we shall prove \eqref{eq.crossing-prop} with appropriately defined trajectories that take into account the effect of a particle's detection on the wave function. In this respect we note, that in Bohmian mechanics the wave function collapse does not happen physically but arises merely as an effective description. It is quite intriguing to see how this effective collapse can be handled very naturally by the so called conditional wave function, the wave function that arises if one or more particles' coordinates are known \cite{duerrobservables:04,duerrteufel:09}. In this respect it is also interesting to investigate the effect of measuring devices on the wave function evolution, for example in an EPR experiment (see e.g.\ \cite{schilling:09}). We have reduced tedious but straightforward technicalities as much as possible to better concentrate on the probabilistic reasoning, which is somewhat new in the context of quantum mechanics.

\subsection{Bohmian mechanics}
\label{subsec.BM}
Bohmian mechanics \cite{bohm:52,duerrteufel:09,duerr:92,duerrobservables:04,holland:95} is a theory of particles in motion that is experimentally equivalent to quantum mechanics whenever the latter makes unambiguous predictions \cite{duerrobservables:04}. The state of $N$  particles is described by their (normalized) wave function $\psi(\bx,t)$, where $\bx=(\bx_1,\,\ldots,\,\bx_N)\in\Rn$, and by their actual configuration (positions) $\bgx=(\bgxone,\,\ldots,\,\bgxN)\in\Rn$. The wave function evolves according to the Schr\"odinger equation
\begin{equation}
\label{eq.schroedinger}
i\hbar\frac{\partial}{\partial t}\psi(\bx,t)=H\psi(\bx,t)
\end{equation}
and governs the motion of the particle by ($l=1,\,\ldots,\,N$)
\begin{equation}
\label{eq.BMeq-of-motion}
\frac{d}{dt}\bbqqtl=\bv_l^\psi\left(\bbqqt,t\right)
      =:\frac{\hbar}{\massl}\Im\left(\frac{\nablal\psi(\bbqqt,t)}{\psi(\bbqqt,t)}\right)\,.
\end{equation}
Here $\bbqqt[0]=\bx_0$ is the particles' configuration at time $t=0$, $m_l$ is the mass of the $l$th particle and $\nablal$ is the gradient with respect to $\bxl$. In \eqref{eq.schroedinger} $H$ is the usual Schr\"odinger Hamiltonian
\begin{equation}
\label{eq.defH}
H=-\sum\limits_{l=1}^{N}\frac{\hbar^2}{2\massl}\Deltal+V(\bx)=H_0+V(\bx)
\end{equation}
with the real valued potential $V$. From now on we shall use natural units $\hbar=m_l=1$.

According to  Born's law, the positions of particles  each having wave function $\psi$ are always $|\psi|^2$-distributed (see \cite{duerr:92} for a precise assertion). The distribution is equivariant: If $\bgx^\psi (\cdot,0)$ is $|\psi(\cdot,0)|^2$-distributed then $\bgx^\psi(\cdot,t)$ is $|\psi(\cdot,t)|^2$-distributed. In terms of the flow map $\Phi^{\psi}_{t_2,t_1}:\Rn\to\Rn$ of \eqref{eq.BMeq-of-motion}, i.e.
\begin{gather}
\notag
\Phi^{\psi}_{t_2,t_1}\left(\bbqqt[t_1]\right)=\bbqqt[t_2]\,,\\
\intertext{this means that}
\label{eq.equiv}
\mathbbppsi[\psi(\cdot,t_2)]=\mathbbppsi[\psi(\cdot,t_1)]\circ\left(\Phi^{\psi}_{t_2,t_1}\right)^{-1}\,.
\end{gather}
For a wide class of sufficiently regular potentials and initial wave functions $\mathbbppsi$-almost sure global existence of Bohmian mechanics was proved in \cite{berndl:95} and \cite{teufeltumulka:05}. In particular, our setting below falls into the scope of Corollary 3.2 in \cite{berndl:95} resp.\,Corollary 4 in \cite{teufeltumulka:05}.


\subsection[Scattering behavior of Bohmian trajectories]{Asymptotic behavior of Bohmian trajectories in scattering situations}
\label{sec.asym-beh}

In scattering situations the Bohmian trajectories become straight lines at large distances from the scattering center. For non-interacting particles,
\begin{equation*}
V(\bx)=\sum\limits_{l=1}^N V_l(\bx_l)\qquad \text{resp.}
      \qquad H=\sum\limits_{l=1}^N H_l=\sum\limits_{l=1}^N\left(-\nhalf[1]\Deltal+V_l(\bx_l)\right)\,,
\end{equation*}
this assertion is a direct extension (cf.\ \cite{roemer:09}) of the corresponding result for one particle \cite{roemer:05} and will be used in the proof of \eqref{eq.crossing-prop}. It holds for a large class of wave functions (see e.g.\cite{duerr1:04,roemer:05} in the one particle case) which we denote by $\mathcal{G}$. The set $\mathcal{G}$ is invariant under the time evolution $e^{-iHt}$ and dense in the set $\bigotimes_{l=1}^N\mathcal{H}_{a.c.}(H_l)$ of ``pure'' scattering wave functions.

\bthe
\label{thm.as-beh}
Let $V$ be sufficiently smooth and fast decaying and let zero be neither a resonance
nor an eigenvalue of $H_l$ ($l=1,\ldots,N$). Let $\psi\in\g$ with $\|\psi\|=1.$ Then:
\begin{enumerate}
\item
The Bohmian trajectories $\bbqqt$ exist uniquely and globally in time for $\mathbbppsi$-almost all initial configurations $\bx_0\in\Rn$.
\item
For $\mathbbppsi$-almost all Bohmian trajectories the asymptotic velocity $\bvinfty(\bx_0):=\lim\limits_{t\to\infty}\frac{\bbqqt}{t}$ exists and
the distribution of $\bvinfty$ under  $\mathbbppsi$ has the  density $|\pouth(\cdot)|^2$.
Moreover for all $\epsilon>0$ there exist $T>0$ and $C>0$ such that
\begin{equation}
\label{eq.P(v-well-behaved)-pure0}
\mathbbppsi\left(\left\{\bx_0\in\Rn\mid
    \left|\bv^\psi\left(\bbqqt,t\right)-\bvinfty(\bx_0)\right|<Ct^{-\frac{1}{2}}\quad
    \forall\,t\geq T\right\}\right)>1-\epsilon\,.
\end{equation}
\end{enumerate}
\ethe
Theorem \ref{thm.as-beh} is based on the asymptotic structure of the wave function $\psi$ at large distances from the scattering center. We note that large distances imply large times. Asymptotically the wave function achieves local plane wave structure, meaning that \eqref{eq.loc-pl-w} holds in a pointwise sense and, moreover, that a corresponding statement holds for gradients. A crucial ingredient in proving these pointwise statements is the expansion in generalized eigenfunctions and the application of stationary phase methods \cite{roemer:09}.

\section{Exit statistics}
\label{sec.ex-stat}
In a realistic scattering experiments when a particle crosses or hits a detector surface some measurement takes place. In $N$-particle scattering each particle is detected at a different time. The measurement of one particle produces a collapse of the entangled wave function and one needs to address the question whether that collapse has an effect on the detection statistics of the as yet not detected particles. To answer this question we shall first consider the mathematically idealized situation, where the detection event is simply modeled by the crossing of a particle's trajectory through a surface without invoking a collapse of the wave function.

\subsection{Exit statistics without collapse}
\label{subsec.ex-stat-without}

The first exit time $\tex[A](\bx_0)$ of the trajectory $\{\bbqqt,\,t\geq0\}$ from an open set $A\subset\Rn[3N]$ is
\begin{equation}
\label{eq.t-exit}
\tex[A](\bx_0):=\inf\left\{t\geq0\mid \bx_0\in A\; \text{ and }\bbqqt\not\in A\right\}\,.
\end{equation}
We are interested in the sets $B_{l,R}:=\{\bx\in\Rn[3N]\mid |\bx_l| = x_l<R\}$. The configuration space trajectory $\{\bbqqt,\,t\geq0\}$ leaves $B_{l,R}$  when the $l$th particle's trajectory leaves the open ball $B_R=\{\bx\in\Rn[3]\mid x<R\}$. Since $\bbqqt$ is continuous $\bbqqtl[\tex(\bx_0)]\in\partial B_R$, the sphere with radius $R$. Let $\Sigma_l\subset\partial B_1$ be a solid angle in $ \mathbb{R}^3$ and $R\Sigma_l$ the corresponding subset of $\partial B_R$. The first exit of the $l$th particle is in $R\Sigma_l$ if and only if $\bbqqtl[\tex(\bx_0)]\in R\Sigma_l$.
With the above we can formulate
\bthe
\label{thm.crossing-prob}
Under the conditions of Theorem 1:
\begin{equation}\label{thm2}
\begin{split}
\lim\limits_{R\to\infty}\mathbbppsi
        \Big(\Big\{\bx_0\in\Rn[3N]\mid\bbqqtl[\tex(\bx)]\in R\Sigma_l\,,\; \forall l=1,2,&\ldots,N\Big\}\Big)\\
                        &=\int\limits_{C_{\Sigma_1}}\hspace{-0.2cm}d^3k_1\;
        \ldots\negthickspace\int\limits_{C_{\Sigma_N}}\hspace{-0.2cm}d^3k_N\;\left|\pouth(\bk)\right|^2\, ,
\end{split}
\end{equation}
where $C_{\Sigma_k}$ denotes the cone in $\mathbb{R}^3$ subtended by the angle $\Sigma_k$.
\ethe

\bpro
By Theorem \ref{thm.as-beh} we only need to show that
\begin{multline}
\label{eq.crossing-prob=vinfty-in-cone}
\lim\limits_{R\to\infty}\mathbbppsi
        \Big(\Big\{\bx_0\in\Rn[3N]\mid\bbqqtl[\tex(\bx)]\in R\Sigma_l\,,\; \forall l=1,2,\ldots,N\Big\}\Big)\\
      =\mathbbppsi\left(\bvinfty\in C_{\Sigma_1}\times C_{\Sigma_2}\times\ldots\times C_{\Sigma_N}\right)\,.
\end{multline}
For this we employ the asymptotic straightness of the Bohmian trajectories that holds for large times. Thinking of $T$ as a large time we split the trajectory into two pieces:
\begin{equation}
\label{splitting}
\begin{split}
\Big|&\bbqqt-\bvinfty(\bx_0) t\Big|\\
 &\leq\begin{cases}
      \sup\limits_{0\leq t\leq T}|\bbqqt|+\vinfty(\bx_0)T&\text{for }0\leq t\leq T\,,\\
      \left|\bbqqt-\bbqqt[T]-\bvinfty(\bx_0)(t-T)\right|+|\bbqqt[T]|+\vinfty(\bx_0)T\,
            &\text{for }t> T\,.
      \end{cases}
\end{split}
\end{equation}
Observe that global existence of the Bohmian trajectories implies that for given $\eps>0$ and  $T$  there exists $C$  such that
\begin{equation}
\label{setA}
\mathbbppsi\Big(\big\{\bx_0\in\Rn\mid\,\sup\limits_{0\leq t\leq T}|\bbqqt|
      <C\big\}\Big)>1-\frac{\epsilon}{3}\,.
\end{equation}
Next, by Theorem 1 (ii) for $K$ large enough
\begin{equation}
\label{setB}
\mathbbppsi\left(\vinfty<K\right)=\int\limits_{B_K}|\pouth(\bk)|^2\,d^{3N}k>1-\frac{\epsilon}{3}\,.
\end{equation}
By integrating the velocities in \eqref{eq.P(v-well-behaved)-pure0}, we obtain the following estimate for the positions: There is $\tilde{C}>0$ and $T>0$ such that
\begin{equation}
\label{setC}
\mathbbppsi\left(\left\{\bx_0\in\Rn\mid
      |\bbqqt-\bbqqt[\tilde{T}]-\bvinfty(\bx)(t-\tilde{T})|
      <\tilde{C}\sqrt{t},\quad\forall t\geq \tilde{T}\right\}\right)>1-\frac{\epsilon}{3}
\end{equation}
for all $\tilde{T}>T$. Observing the splitting \eqref{splitting} we may thus conclude that there exists a $C_{\eps}$ such that
\begin{equation}
\label{eq.P(v-well-behaved)-pure}
\mathbbppsi\left(\left\{\bx_0\in\Rn\big|\,
    \big|\bbqqt-\bvinfty(\bx_0)t\big|<C_\eps(1+\sqrt{t}),\quad\forall t\geq 0\right\}\right)>1-\epsilon\,.
\end{equation}
We set
$$
G_C:=\left\{\bx_0\in\Rn\big |\;\big|\bbqqt-\bvinfty(\bx_0)t\big|<C(1+\sqrt{t})\,,\;\forall t\geq 0\right\}.
$$
and call the trajectories on this set asymptotically straight. Note that we have just shown that typical\linebreak
\vspace{-0.62cm}
\begin{floatingfigure}[l]{7.5cm}
\vspace{-0.5cm}
\begin{center}
\includegraphics[width=6cm]{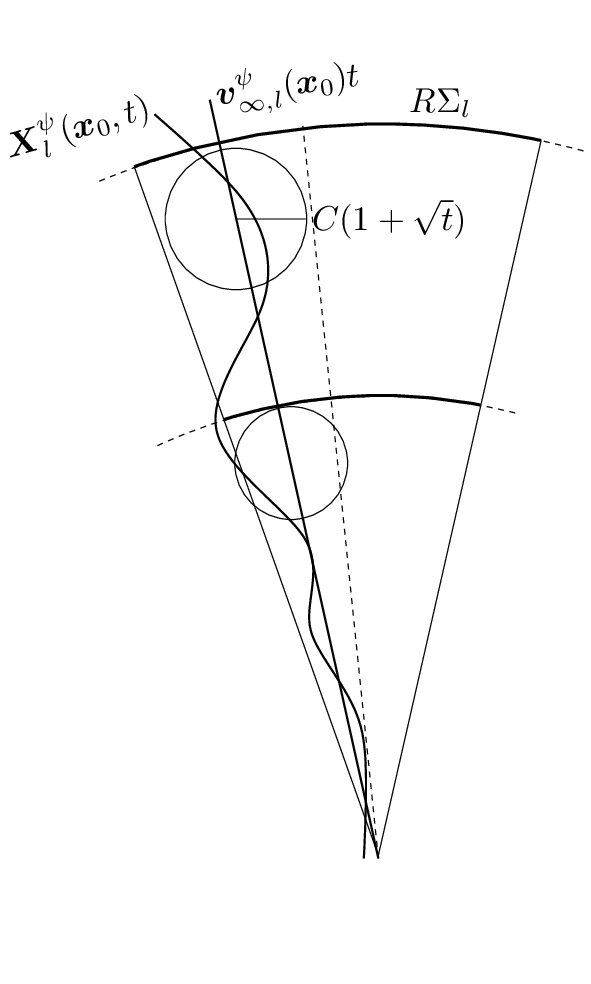}
\end{center}
\vspace{-1.2cm}
\caption{Real (asymptotically straight) and straight trajectory.}
\label{figtraj}
\end{floatingfigure}

trajectories are asymptotically straight. Since the distribution of $\bvinfty$ is absolutely continuous we may assume without loss of generality that $0\neq\bvinfty$ and that the cones $C_{\Sigma_k}$ are open.  Let $\bvinfty(\bx_0)\in C_{\Sigma_1}\times\ldots\times C_{\Sigma_N}$. Then the straight trajectory $\bvlinfty(\bx_0)t$ of the $l$th particle crosses the surface $R\Sigma_l$ at time $T_R:=R/\vlinfty(\bx_0)$. Thus the crossing time of the straight trajectory grows linearly with the distance $R$. The same is true for the distance between the point where $\bvlinfty(\bx_0)t$ crosses $R\Sigma_l$ and the boundary of the surface $R\Sigma_l$. However, since the difference between the corresponding asymptotically straight trajectory $\bbqqtl$ and the straight trajectory $\bvlinfty(\bx_0)t$ grows only sublinearly in time ($\sim \sqrt{t}$), this difference evaluated at $T_R$ also grows sublinearly with $R$. Hence, if $R$ is big enough, at $T_R$ the distance between the point of crossing of the straight trajectory and the boundary of the surface $R\Sigma_l$ is larger than the distance between $\bbqqtl[T_R]$ and $\bvlinfty(\bx_0)T_R$. Since the trajectory is continuous in $t$ this implies that $\bbqqtl$ crosses $\partial B_R$ first in $R\Sigma_l$, if and only if $\bvlinfty(\bx_0)$ lies in $C_{\Sigma_l}$ (see Figure \ref{figtraj}).
\epro


\subsection{Exit statistics with collapse}
\label{subsec.ex-stat-with}
The actual detection of a  particle changes the wave function; it collapses, so that the new spatial support
of the wave function is in accordance with  the measured position. In standard quantum mechanics such a  collapse is introduced when the particle is measured at a given fixed time $t$. In the situation at hand however the detection time is random and the collapse must be invoked at that random time.  In Bohmian mechanics the collapse of the wave function does not actually happen. The collapsed wave function is replaced by the well defined concept of conditional wave function which is based on the existence of trajectories and can be used to describe the dynamics of subsystems \cite{duerr:92,duerrteufel:09,duerrobservables:04}. For example, -- \emph{given the first particle's trajectory $\bbqqtone$} -- the conditional wave function for the remaining $N-1$ particles is
\begin{equation*}
\psi^\text{cond}(\bx_2,\ldots,\bx_N, t):=\frac{\psi\big(\bbqqtone,\bx_2,\ldots,\bx_N,t\big)}{\big\|\psi\big(\bbqqtone,\cdot,\ldots\cdot,t\big)\big\|}\,.
\end{equation*}
The conditional wave function $\psi^\text{cond}$ is random since the first particle's trajectory $\bbqqtone$ is random. It yields the conditional probability $\PP\left(\big(\bbqqttwo,\ldots,\bbqqtN\big)\in\cdot \mid\bbqqtone=\bx_1\right)=\PP[\psi^\text{cond}(\cdot,t)](\cdot)$: a straightforward calculation shows that
\begin{equation*}
\begin{split}
\PP\Big(\big(\bbqqttwo,&\ldots,\bbqqtN\big)\in d^3x_1\ldots d^3x_N \mid\bbqqtone=\bx_1\Big)\\
      &=\frac{|\psi(\bx_1,\bx_2,\ldots,\bx_N,t)|^2}{\|\psi(\bx_1,\cdot,\ldots,\cdot,t)\|^2}d^3x_1\ldots d^3x_N
      	=:|\psi_{\bx_1,t}(\bx_2,\ldots,\bx_N)|^2d^3x_1\ldots d^3x_N\\
      &\hspace{8cm}=\PP[\psi^\text{cond}(\cdot,t)]\left(d^3x_1\ldots d^3x_N\right)\,.
\end{split}
\end{equation*}

In measurement-like situations the conditional wave function becomes the so called effective wave function which coincides with the collapsed wave function of standard quantum mechanics \cite{duerr:92,duerrteufel:09,duerrobservables:04}.

Now let
\begin{equation}
\label{eq.def-TR}
\TR:=\min\{\texone,\,\textwo\,\dots, t_{\rm ex}^{B_{N,R}}(\bx_0)\}
\end{equation}
be the time of first detection. Without loss of generality we assume that particle one is detected first, $\TR = \texone$. Then its measured position is $\bbqqtone[\TR]$ and the remaining particles' conditional wave function at time $\TR$ is
\begin{equation}
\label{condwf1}
\psi^\text{cond}(\bx_2,\ldots,\bx_N,\TR)
      =\frac{\psi\big(\bbqqtone[\TR],\bx_2,\ldots,\bx_N,\TR\big)}
      {\big\|\psi\big(\bbqqtone[\TR],\cdot,\ldots,\cdot,\TR\big)\big\|}\,.
\end{equation}
To ease the discussion we assume that after its detection the particle is stuck at its measured position. Other
more complicated scenarios where the detected particle is processed within the detector do not add further insight
into the question at hand: eventually they would lead to the same technicalities as presented here. Thus the conditional wave function's time evolution after time $\TR$ is given by
\begin{equation}
\label{condwf2}
\psi^\text{cond}(\bx_2,\ldots,\bx_N,t)=e^{-i(H_2+\cdots+ H_N)(t-\TR)}
      \frac{\psi\big(\bbqqtone[\TR],\bx_2,\ldots,\bx_N,\TR\big)}
      {\big\|\psi\big(\bbqqtone[\TR],\cdot,\ldots,\cdot,\TR\big)\big\|}\,.
\end{equation}
This time evolved conditional wave function then defines the evolution of the remaining $(N-1)$ particles after time $\TR$:
\begin{equation}
\frac{d}{dt}\bgx^{\psi^\text{cond}}_l(\bx_0,t)
	=\bv_l^{\psi^\text{cond}}\big(\bgx^{\psi^\text{cond}}_2(\bx_0,t),\ldots,\bgx^{\psi^\text{cond}}_N(\bx_0,t),t\big),
	\qquad l=2,\dots,N,\;t\geq \TR\,,
\end{equation}
with ``initial" conditions $\bgx^{\psi^\text{cond}}_l(\bx_0,\TR)=\bbqqtl[\TR]$.

On the set of initial configurations $\bx_0$ such that $\TR=\texone$ one is thus led to the stopped process
\begin{equation}
\label{eq.def-measur-wf}
\bgy^R(\bx_0,t):= \begin{cases}
                  \bgx^{\psi}(\bx_0,t) &\text{for } t<\TR\\
                  \left(\bbqqtone[\TR],\bgx_2^{\psi^\text{cond}}(\bx_0,t),\ldots,
                  \bgx_N^{\psi^\text{cond}}(\bx_0,t)\right)  &\text{for } t\geq\TR.
                  \end{cases}
\end{equation}
In the following for ease of notation we denote the measure $\PP$ conditioned on $\TR=\texone$ again by $\PP$.
It is not a priori clear that the stopped process $(\bgy^R(\bx_0,t))_{t \geq 0}$ exists globally and uniquely for $\mathbb{P}^{\psi}$-almost every $\bx_0$. That this is the case is the new technical ingredient in this section.
\bthe
\label{thm.measured-as-beh}
Under the conditions and notations of Theorem \ref{thm.as-beh} the following holds.
\begin{enumerate}
\item
For every $R>0$ the stopped process $(\bgy^R(\bx_0,t))_{t \geq 0}$ exists globally and uniquely for $\mathbb{P}^{\psi}$-almost every $\bx_0$.
\item
Let $R>0$, $T>0$ and $C>0$. We define the set $\V$ of initial values $\bx_0$, which fulfill
\begin{equation}\label{eq.as-vel-before-collapse}
\forall\;\TR >t\geq T:\quad\Big|\bv^\psi\left(\bgy^R(\bx_0,t),t\right)-\bvinfty(\bx_0)\Big|\leq \frac{C}{\sqrt{t}}
\end{equation}
and
\begin{equation}\label{eq.as-vel-after-collapse}
\forall\;t\geq \TR,\;l \neq 1:\quad\left|\bv_l^{\psi^{cond}}\left(\bgy^R(\bx_0,t),t\right)-\bvlinfty(\bx_0)\right|\leq \frac{C}{\sqrt{\TR}}.
\end{equation}
Then for all $\eps>0$ there exist $T_{\eps}>0,\,C_{\eps}>0$ and $R_{\eps}>0$ such that for all $T \geq T_{\eps}$, $C \geq \,C_{\eps}>0$ and $R \geq R_{\eps}>0$
\begin{equation}
\label{main}
\mathbbppsi(\V)>1-\eps.
\end{equation}
\end{enumerate}
\ethe
We remark that this theorem is analog to Theorem \ref{thm.as-beh}. The only difference is that the error term between the conditional wave function and the asymptotic local plane wave (cf.\ \eqref{eq.loc-pl-w}) and thus the error term between the ``real" velocity $\bv_l^{\psi^\text{cond}}$ and the asymptotic velocity $\bvlinfty$ is now dominated by the first detection time $\TR$. Thus, once global existence and uniqueness of the stopped process, i.e.\ assertion (i), is established, the proof of assertion (ii) is completely analog to the proof of assertion (ii) of Theorem \ref{thm.as-beh} (cf.\ \cite{roemer:09}).

{\bf Proof of assertion (i):}\\
The global existence and uniqueness of Bohmian trajectories was proven in \cite{berndl:95} under the condition that the wave function $\psi$ belongs to $C^{\infty}(H)=\bigcap_{n \in \mathbb{N}}\mathcal{D}(H^n)$, where $\mathcal{D}(H^n)$ denotes the domain of $H^n$. Hence we need to show that $\psi^{cond}$ belongs to $C^{\infty}(\Htild)$ where $\Htild:=\sum_{i=2}^N H_i$. Omitting the normalization factor in \eqref{condwf1}, this amounts to proving that $\psi(\bgx_1,\cdot)\in\mathcal{D}(\Htild^n)$ for every $\bgx_1\in\Rn[3]$ and $n\in\N$. Set $\phi_n(\bgx_1,\cdot):=\Htild^n\psi(\bgx_1,\cdot)$. Since $\psi\in C^{\infty}(H)$ and thus $\int\limits_{\Rn[3(N-1)]}\|(H_1\phi_n)(\cdot,\bz)\|^2d^{3(N-1)}z=\|H_1\phi_n\|^2\leq\|H^{n+1}\psi\|^2<\infty$, this implies $\|(H_1\phi_n)(\cdot,\bz)\|<\infty$ for almost every (with respect to Lesbesgue measure) $\bz\in\Rn[3]$ and thus
\begin{equation*}
\phi_n(\cdot,\bz)\in\mathcal{D}(H_1)=W^2(\Rn[3])\qquad\text{for a.\,e. }\bz\in\Rn[3(N-1)]\,.
\end{equation*}
Here $W^2(\Rn[3])$ is the second Sobolev space. Thus we can apply an instance of the Gagliardo-Nirenberg inequality \cite{gagliardo:59,nirenberg:59}, namely
\begin{equation*}
\|u\|_{L^\infty(\Rn[3])}\leq C\|D^2u\|^\frac{3}{4}_{\ltrN[3]}\|u\|^\frac{1}{4}_{\ltrN[3]}
\end{equation*}
with $\|D^mu\|_{\ltrN[3]}:=\max\limits_{|\alpha|=m}\|D^\alpha u\|_{\ltrN[3]}$ and $C>0$ independent of $u\in W^2(\Rn[3])$, to get
\begin{equation*}
\|\phi_n(\cdot,\bz)\|_{L^\infty(\Rn[3])}
      \leq C\|D^2_{x_1}\phi_n(\cdot,\bz)\|^\frac{3}{4}_{L^2(\Rn[3])}
      \|\phi_n(\cdot,\bz)\|^\frac{1}{4}_{L^2(\Rn[3])}\qquad\text{for a.\,e. }\bz\in\Rn[3(N-1)]\,.
\end{equation*}
Then, using H\"older in the second to last step,
\begin{align*}
\|\phi_n(\bgx_1,\cdot)\|&^2_{L^2(\Rn[3(N-1)])}
            =\int\limits_{\Rn[3(N-1)]}|\phi_n(\bgx_1,\bz)|^2d^{3(N-1)}z
            \leq\int\limits_{\Rn[3(N-1)]}\|\phi_n(\cdot,\bz)\|^2_{L^\infty(\Rn[3])}d^{3(N-1)}z\\
      &\leq C^2\int\limits_{\Rn[3(N-1)]}\|D^2_{x_1}\phi_n(\cdot,\bz)\|^\frac{3}{2}_{L^2(\Rn[3])}
            \|\phi_n(\cdot,\bz)\|^\frac{1}{2}_{L^2(\Rn[3])}d^{3(N-1)}z\\
      &\leq C^2\left[\int\limits_{\Rn[3(N-1)]}
            \|D^2_{x_1}\phi_n(\cdot,\bz)\|^{\frac{3}{2}\cdot\frac{4}{3}}_{L^2(\Rn[3])}d^{3(N-1)}z\right]^{\frac{3}{4}}
            \left[\int\limits_{\Rn[3(N-1)]}\|\phi_n(\cdot,\bz)\|^{\frac{1}{2}\cdot 4}_{L^2(\Rn[3])}
            d^{3(N-1)}z\right]^{\frac{1}{4}}\\
      &\leq C^2\|D^2\phi_n\|^\frac{3}{2}_{L^2(\Rn[3N])}\|\phi_n\|^\frac{1}{2}_{L^2(\Rn[3N])}
\end{align*}
\emph{for every} $\bgx_1\in\Rn[3]$. Moreover, since $\|H\phi_n\|\leq\|H^{n+1}\psi\|<\infty$, i.e.\ $\phi_n\in\mathcal{D}(H)=\mathcal{D}(H_0)=W^2(\Rn[3N])$, the last term is finite, i.e.\ we have just shown that indeed $\phi_n(\bgx_1,\cdot)\in L^2(\Rn[3(N-1)])$ for arbitrary  $\bgx_1\in\Rn[3]$ and $n\in\N$. This proves that $\psi^{cond}$ belongs to $C^{\infty}\left(\Htild\right)$.

The second step consists in proving that the dynamics of the stopped process for times $t>\TR$ is well defined, i.e.\ that for $\mathbb{P}^{\psi}$-almost all initial configurations $\bx_0$ the random vector
\begin{displaymath}
\big(\bbqqttwo[\TR],\cdots,\bbqqtN[\TR]\big)
\end{displaymath}
belongs to the set of initial configurations $\bz_0$, for which
\begin{equation}
\begin{split}
\frac{d}{d\tau}\bgz^{\psi^{cond}}(\bz_0,\tau)&=\bv^{\psi^{cond}}(\bgz^{\psi^{cond}}(\bz_0,\tau),\tau)\,,\\
      \bgz^{\psi^{cond}}(\bz_0,0)&=\bz_0
\end{split}
\end{equation}
has unique global solutions. Let $\G^{\psi}$ be the set of initial conditions, for which the flow map $\flow[t_2,t_1]$ of \eqref{eq.BMeq-of-motion} exists globally. In \cite{berndl:95} is was shown that $\mathbb{P}^{\psi}(\G^{\psi})=1$. Set
\begin{equation*}
G^R:=\left\{\bx_0\in \G^\psi\mid \big(\bbqqttwo[\TR],\cdots,\bbqqtN[\TR]\big) \in \G^{\psi^{cond}}\right\}\,.
\end{equation*}
We wish to show that $\mathbbppsi(G^R)=1\,.$ By the definition of conditional probability
\begin{equation}
\label{eq.P(G^R_2)=int(cond-prob)}
\begin{split}
\PP&(G^R)=\hspace{-0.5cm}\int\limits_{\Rn[3]\times[0,\infty)}\hspace{-0.5cm}\mathbbppsi\big(G^R\mid
            \big(\bgxpsi_1(\cdot,T^R),T^R\big)=(\bx_1,t)\big)\mathbbppsi_{(\bgxpsi_1(\cdot,T^R),T^R)}(d^3x_1dt)\\
      &=\hspace{-0.5cm}\int\limits_{\Rn[3]\times[0,\infty)}\hspace{-0.5cm}\mathbbppsi
            \big(\big(\bgxpsi_2(\cdot,t),\ldots,\bgxpsi_N(\cdot,t)\big)\in\G^{\psixt[\bx_1,t]}\mid
            \big(\bgxpsi_1(\cdot,t),T^R\big)=(\bx_1,t)\big)\mathbbppsi_{(\bgxpsi_1(\cdot,T^R),T^R)}(d^3x_1dt)
\end{split}
\end{equation}
where $\mathbbppsi_{(\bgxpsi_1(\cdot,T^R),T^R)}(d^3x_1dt)$ denotes the image measure of the random vector $(\bgxpsi_1(\cdot,T^R),T^R)$.
Recall the flow map $\flow[t_2,t_1]$ of \eqref{eq.BMeq-of-motion}. Denoting by $\flow[0,t]$ the inverse of $\flow[t,0]$ and by $\bgx_l$ the random variables $\bgx_l(\bx)=\bx_l$ ($l=1,\ldots,N$), we have by equivariance that
\begin{equation}
\label{eq.transported-cond-prob}
\begin{split}
\PP\big((\bgxpsi_2(\cdot,t),\dots, \bgxpsi_N(\cdot,t))\in&\G^{\psixt[\bx_1,t]}
            \mid(\bgxpsi_1(\cdot,T^R),T^R)=(\bx_1,t)\big)\\
	&=\PP[\psi(\cdot,t)]\big((\bgx_2,\ldots,\bgx_N)\in\G^{\psixt[\bx_1,t]}
            \mid (\bgxpsi_1(\cdot,T^R),T^R)\circ\flow[0,t]= (\bx_1,t)\big)
\end{split}
\end{equation}
for all $t\in\R$. Introducing this into \eqref{eq.P(G^R_2)=int(cond-prob)} gives
\begin{equation}
\label{aftertimeshift}
\PP(\G^R)\negthickspace=\hspace{-0.6cm}\int\limits_{\Rn[3]\times[0,\infty)}\hspace{-0.5cm}
	\PP[\psi(\cdot,t)]\big((\bgx_2,\ldots,\bgx_N)\in\G^{\psixt[\bx_1,t]}
            \mid (\bgxpsi_1(\cdot,T^R),T^R)\circ\flow[0,t]=(\bx_1,t)\big)\PP_{(\bgxpsi_1(\cdot,T^R),T^R)}(d^3x_1dt)\,.
\end{equation}
Next we observe that
\begin{equation}
\label{finalequ}
\begin{split}
\PP[\psi(\cdot,t)]\big((\bgx_2,\ldots,\bgx_N)\in\G^{\psixt[\bx_1,t]}
            \mid (\bgxpsi_1(\cdot,T^R),T^R)\circ\flow[0,t]= &(\bx_1,t)\big)\\
	&=\PP[{\psix[\bx_1,t]}]\left(\G^{\psixt[\bx_1,t]} \mid T^R\circ\flow[0,t] = t\right) =1
\end{split}
\end{equation}
where the second equality is a direct consequence of $\mathbbppsi[{\psix[\bx_1,t]}]\left(\G^{\psixt[\bx_1,t]}\right)=1$. Introducing this in \eqref{aftertimeshift} we get the desired almost sure global existence $\mathbbppsi(G^R)=1\,.$
To see the first equality in \eqref{finalequ} we use
\begin{equation*}
\PP[\psi(\cdot,t)]_{(\bgx_1,T^R\circ\flow[0,t])}(d^3x_1ds)
      =\|\psi(\bx_1,\cdot,t)\|^2\PP[\psi_{\bx_1,t}]_{T^R\circ\flow[0,t](\bx_1,\cdot)}(ds)d^3x_1
\end{equation*}
and compute: Let $D \subset \mathbb{R}^3 \times \Rn[3(N-1)]$ and split $\bx=(\bx_1,\bz)$. Then
\begin{align*}
\PP[\psi(\cdot,t)]\big(\bgz\in\G^{\psixt[\bx_1,t]},&(\bgx_1,T^R\circ\flow[0,t])\in D\big)
	      =\hspace{-1.4cm}\int\limits_{\{\bx\mid\bz\in \G^{\psixt[\bx_1,t]},(\bx_1,T^R\circ\flow[0,t](\bx))\in D\}}
            \hspace{-1.4cm}|\psi(\bx,t)|^2d^{3N}x\\
	&=\int\limits_{\Rn[3]}\Big(\|\psi(\bx_1,\cdot,t)\|^2\hspace{-1.4cm}
            \int\limits_{\{\bz\in \G^{\psixt[\bx_1,t]}\mid(\bx_1,T^R\circ\flow[0,t](\bx_1,\bz))\in D\}}\hspace{-1.5cm}
    	      |\psixt[\bx_1,t](\bz)|^2d^{3(N-1)}z\Big)d^3x_1\\
	&=\int\limits_{\Rn[3]}\|\psi(\bx_1,\cdot,t)\|^2\,
		\PP[{\psixt[\bx_1,t]}]\big(\G^{\psixt[\bx_1,t]}, T^R\circ\flow[0,t](\bx_1,\cdot)\in
            \{s\mid(\bx_1,s)\in D\}\big)d^3x_1\\
	&=\int\limits_{\Rn[3]}\Big(\|\psi(\bx_1,\cdot,t)\|^2\hspace{-0.5cm}
		\int\limits_{\{s\mid(\bx_1,s)\in D\}}\hspace{-0.5cm}\PP[{\psixt[\bx_1,t]}](\G^{\psixt[\bx_1,t]}\mid
		T^R\circ\flow[0,t](\bx_1,\cdot)=s)\PP[{\psixt[\bx_1,t]}]_{T^R\circ\flow[0,t](\bx_1,\cdot)}(ds)\Big)d^3x_1\\
	&=\int\limits_D\PP[{\psixt[\bx_1,t]}](\G^{\psixt[\bx_1,t]}\mid  T^R\circ\flow[0,t] = s)
            \PP[\psi(\cdot,t)]_{(\bgx_1,T^R\circ\flow[0,t])}(d^3x_1ds)\,,
\end{align*}
i.e.\ by the definition of the conditional probability
\begin{equation*}
\PP[\psi(\cdot,t)]\left(\bgz\in\G^{\psixt[\bx_1,t]}\mid (\bgxpsi_1(\cdot,T^R),T^R)\circ\flow[0,t]= (\bx_1,s)\right)
	=\PP[{\psix[\bx_1,t]}]\left(\G^{\psixt[\bx_1,t]} \mid T^R\circ\flow[0,t] = s\right)
\end{equation*}
for all $s,t\in\R$. Setting $s=t$ gives the first equality in \eqref{finalequ}.
\section{Conclusion}
It is of course desirable to extend the results to $N$-particle scattering, i.e.\ to the case of interacting particles. However, our analysis here relies heavily on earlier works, where it was shown that the scattered wave function attains local plane wave structure \eqref{eq.loc-pl-w} with a small pointwise error. This was achieved using an expansion into generalized eigenfunctions, for which regularity properties had been established in \cite{duerr1:04,teufel2:99}. To apply this method to interacting particles we would need similar regularity properties for the generalized eigenfunctions also in this case. These are however not known and presumably much more difficult to get.



\providecommand{\bysame}{\leavevmode\hbox to3em{\hrulefill}\thinspace}
\providecommand{\MR}{\relax\ifhmode\unskip\space\fi MR }
\providecommand{\MRhref}[2]{%
  \href{http://www.ams.org/mathscinet-getitem?mr=#1}{#2}
}
\providecommand{\href}[2]{#2}

\end{document}